\documentclass[twocolumn,
amsmath,
amssymb, 
prl]{revtex4}

\usepackage{graphicx}
\usepackage{subfigure}
\usepackage{dcolumn}
\usepackage{bm}
\usepackage{ulem}
\usepackage{color}

\begin{document}
\normalem

\title{Thermodynamic equilibrium conditions of graphene films on SiC}

\author{Lydia Nemec}
\author{Volker Blum}
\author{Patrick Rinke}
\author{Matthias Scheffler}
\affiliation{Fritz-Haber-Institut der Max-Planck-Gesellschaft, D-14195, Berlin, Germany}

\date{\today}

\begin{abstract}
First-principles surface phase diagrams reveal that epitaxial
monolayer graphene films on the Si side of 3C-SiC(111) can exist as
thermodynamically stable phases in a narrow range of experimentally
controllable conditions, defining a path to the
highest-quality graphene films. Our calculations are
based on a van der Waals corrected density functional. The
full, experimentally observed ($6 \sqrt{3}\times 6
\sqrt{3}$)-R30$^\circ$ supercells for zero- to trilayer
graphene are essential to describe the correct interface
geometries and the relative stability of surface phases and possible
defects.
\end{abstract}

\maketitle

The growth of wafer-size graphene films on a semiconducting substrate
is a first step towards graphene based electronics. The semiconductor SiC as a
substrate may hold the key to device applications. Here, well-ordered
graphene films can be grown directly on a semiconducting substrate by a simple
process (Si sublimation from the
surface, e.g.,
Refs. \cite{bommel1975laa,forbeaux1998hgo,emtsev2009tws,riedl2010rev,heer2011laa,srivastava2012iso,yazdi2013gol})
and the standard tools of  
semiconductor technology can be used for further manipulation.
Indeed, graphene-based devices and even integrated circuits
\cite{lin2011wsg,her2012ttg} were already created on the Si side of
SiC substrates. However, controlling the 
precise thickness of graphene films is important to minimize the
coexistence of monolayer graphene (MLG) and bilayer graphene (BLG).\cite{berger2004ueg,
  emtsev2009tws, first2010ego, heer2011laa} MLG
areas exhibit no band gap, while BLG areas do.\cite{riedl2010rev}
A recent, joint experimental-theoretical study finds particularly high local
resistances across monolayer-bilayer graphene junctions on the
same surface,\cite{ji2011ast} a possible contributing factor to low
carrier mobilities in graphene on
SiC(0001).\cite{emtsev2009tws,ji2011ast} 

The growth of graphene on SiC is special in the sense that, instead of offering one or
more of the components from the gas phase, graphene areas are formed by controlled
sublimation of Si from the
surface.\cite{bommel1975laa,forbeaux1998hgo} Graphene films grown
under ultrahigh vacuum
conditions are typically
inhomogeneous.\cite{emtsev2009tws,heer2011laa} To improve
their quality is therefore a major and ongoing experimental goal.\cite{emtsev2009tws,heer2011laa}
In the past, the appearance of different phases was often interpreted
\cite{emtsev2008iga,ming2011mft,heer2011laa} as intuitive, successive intermediates formed by an
outgoing Si flux that ultimately leads to bulk-like
graphite layers. If graphene films of various
thicknesses were a result of a purely kinetically limited
Si sublimation process (controlled by growth temperature and time), improving upon the homogeneity
of the layer thickness would be difficult.
It would be helpful if there were a set of \emph{thermodynamic} equilibrium conditions
(e.g., temperature $T$ and partial pressures $p$ of Si and C \cite{reuter2002csa}) at
which a desired film thickness were thermodynamically stable over all
others. As long as $T$ is sufficiently high to overcome the relevant
kinetic barriers, large-scale ordered films of that particular
thickness could then be formed simply by finding and maintaining ($p,T$)
near these equilibrium conditions. 

In fact, experimental evidence is mounting that controlled $(p,T)$
conditions do indeed aid the growth of graphene on
SiC. Ref. \cite{tromp2009tak} demonstrates that the carbon-rich
``zero-layer graphene'' (ZLG) or ``buffer layer'' precursor phase 
\cite{bommel1975laa,forbeaux1998hgo,riedl2007spo,emtsev2008iga} (not
yet graphene) on the Si face is a reversible thermodynamic equilibrium
phase at high $T$ with a controlled disilane background
pressure. Reversibility is much harder to demonstrate once a complete
graphene plane is formed,\cite{Hannonprivate} but an increased growth
temperature in an Ar background buffer gas 
does lead to much improved MLG film homogeneity.\cite{emtsev2009tws} 
Excellent wafer-size MLG films are also reported for growth in a confined
cavity that may retain a finite, well-defined Si background pressure
as Si evaporates from the surface.\cite{heer2011laa} Finally, a
well-defined graphene precursor phase on the C face at finite disilane 
background pressure was reported very recently.\cite{srivastava2012iso}
What is still not clear, however, is whether MLG itself is an
equilibrium phase under certain conditions. If so, one could ideally
facilitate the growth of MLG but not BLG on SiC(111). 

We here present first-principles theoretical evidence that such equilibrium
conditions indeed exist for MLG (and, possibly, even BLG) on the Si
face of SiC(111). We employ
density-functional theory 
(DFT) using the van der Waals (vdW) corrected \cite{tkatchenko2009amv}
PBE density functional\cite{perdew1997gga} (called PBE+vdW throughout
this work). Strategies to include vdW effects have become an active
scientific area of their own (see, e.g.,
Refs. \cite{Grimme2011,vdWDF,lee2010hav,Angelos,Zhang2011,mbd} and many  
references therein). However, for the large, 
complex carbon-rich interfaces of interest here, no unambiguously
quantitatively improved approach over the level of theory used here
exists to our knowledge, i.e., the present incorporation of vdW terms
reflects the state of the art. For the structure of the
relevant bulk phases, the impact of different standard
functionals is well understood and systematic. Predicted lattice
parameters (see supporting information (SI) for full reference
data\cite{supplementary})  
for 3C SiC are within 1~\% of PBE+vdW and experiment:
$a_\text{3C-SiC}=$4.33~{\AA},  
4.38~{\AA}, 4.36~{\AA}, respectively, 
for the local density approximation (LDA), PBE without 
vdW, and PBE+vdW. Similary small discrepancies arise for diamond C 
and for the in-plane lattice parameter of graphite:
$a_\text{graphite}$=2.45~{\AA} (LDA), 2.47~{\AA} (PBE), 
2.46~{\AA} (PBE+vdW). Zero-point corrections (ZPC) to the lattice
parameters are also below 1~\%.\cite{supplementary}. For the interplanar lattice 
parameter $c$ of graphite, vdW effects must be included into the PBE
functional (LDA: 6.65~{\AA}, PBE+vdW: 6.66~{\AA}, but 8.65~{\AA} for
PBE). On a technical level, our calculations are based on the FHI-aims
all-electron code\cite{blum2009aim,havu2009eoi} with ``tight''
numerical settings and the massively parallel ELPA eigensolver
library\cite{auckenthaler2011pso} to guarantee accurate total energies
for the very large structure sizes involved (see 
SI for details \cite{supplementary}).

In experiment, one encounters a series of phases on the Si side of
3C-SiC(111) (experimentally also observed on the Si side of
4H-SiC(0001) and 6H-SiC(0001)) when going from a 
Si-rich to a C-rich regime. In UHV, a (3$\times$3) Si-rich layer \cite{starke1998nrm} can
be prepared. Upon Si sublimation, a simpler ($\sqrt{3}\times\sqrt{3}$)-R30$^\circ$
bulk-terminated surface with one adsorbed Si adatom per three unit cells follows.
\cite{kaplan1989ssa, starke1999stf} 
Removing yet more Si creates a partially covalently bonded carbon ``buffer
layer,'' the ZLG phase, with a large, commensurate unit cell:
One (13$\times$13) honeycomb graphene-like supercell (338 C atoms) on a
(6$\sqrt{3}\times$6$\sqrt{3}$)-R30$^\circ$ mesh (108 Si and 108 C atoms per bilayer) of the underlying SiC
substrate.\cite{bommel1975laa} Compared to a graphene plane in graphite, the lattice match
is almost strain-free (experiment: 0.2~\% at $T$=0~K
\cite{li1986teo,baskin1955}, PBE+vdW: 0.1~\%~
\cite{supplementary}). The ZLG phase does not yet exhibit the
electronic properties of actual freestanding 
graphene.\cite{emtsev2008iga} Further heating detaches the ZLG C plane from the substrate
to form MLG and a new C-rich layer underneath.\cite{seyller2008ega,hannon2011dmo} The same
process can be continued to successively form BLG \cite{ohta2006cte} and multilayer
graphene films. Importantly,
our calculations address
the graphene-like films in their experimentally observed, large commensurate
($6 \sqrt{3} \times 6 \sqrt{3}$)-R30$^\circ$ supercells, using slabs containing six
Si-C bilayers under each reconstructed phase (1,742 up to 2,756 atoms for ZLG up
to three-layer graphene (3LG), respectively). The top three SiC bilayers and all
planes above are fully relaxed (residual energy gradients: 8$\cdot$10$^{-3}$~eV/{\AA} or
below). 

\begin{figure}
  \includegraphics[width=0.5\textwidth]{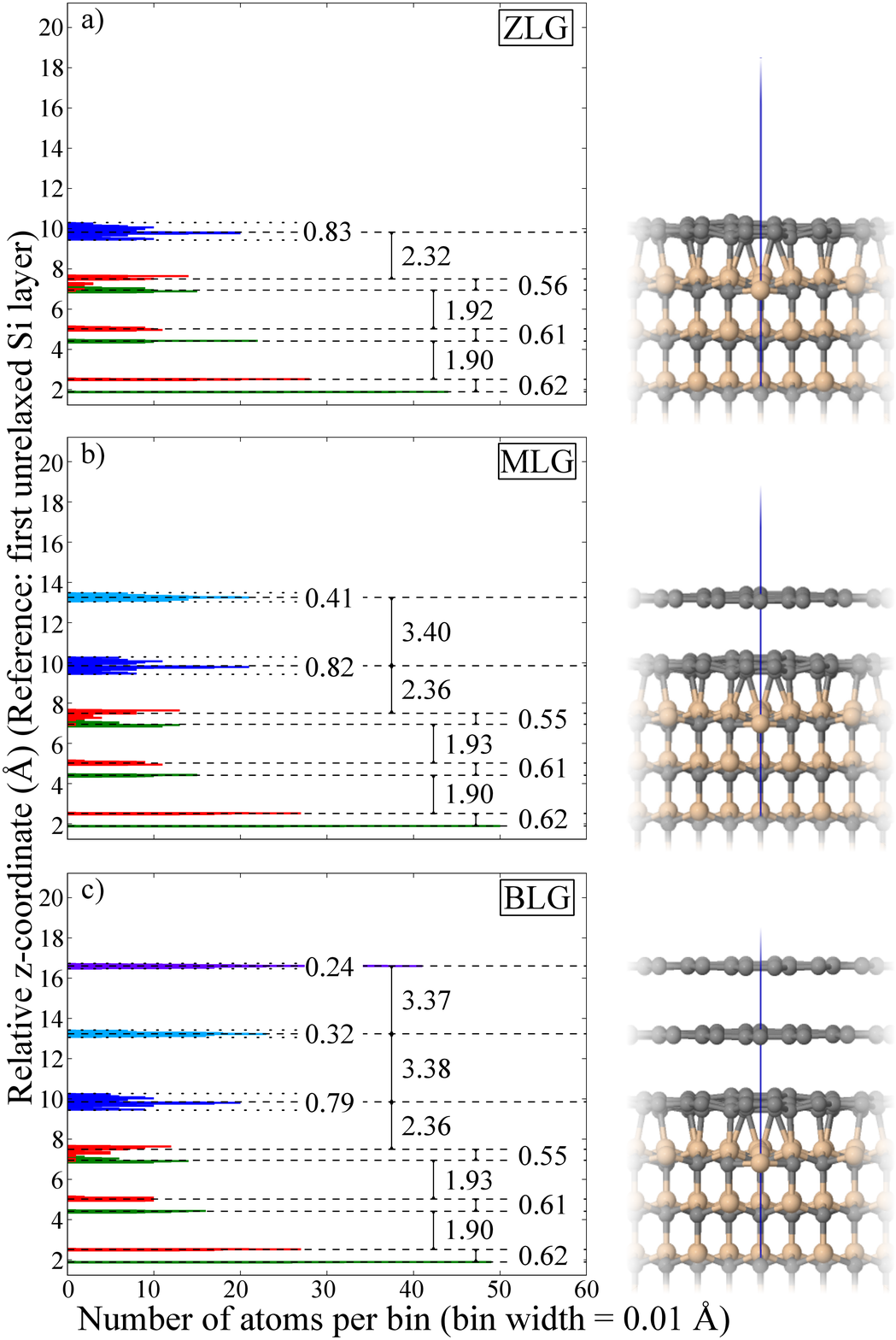}
  \caption{ a-c: Geometry and key geometric parameters for the three phases ZLG,
MLG and BLG on the Si face of 3C-SiC(111), as determined by DFT-PBE+vdW. The histograms
distribute the atoms at each $z$ coordinate (relative to the fourth bilayer of the
substrate) into bins of 0.01~{\AA}~width, giving an impression of the overall corrugation in
each plane.}
\label{Fig:structure}
\end{figure}

Figures \ref{Fig:structure}a-c show the ZLG, MLG and BLG phases together
with key geometry parameters predicted at the level of PBE+vdW. Since
all planes are corrugated,\cite{kim2008ooa,varchon2008rie,sclauzero2012cra} 
histograms for the atomic $z$ coordinates are included. The interface geometry stays
essentially the same despite the addition of more graphene planes. In the MLG phase, we
see a significant buckling of the topmost graphene layer ($\approx$0.41~{\AA} between top
and bottom of the plane). This strong buckling is qualitatively consistent with existing
STM images \cite{chen2005aso,deheer2010ego,berger2010epo}. In the BLG phase, the corrugation is slightly
reduced, but the two top planes are still buckled by 0.24~{\AA} and 0.32~{\AA},
respectively. This buckling reflects some coupling to the covalently
bonded interface C-rich plane, which is much more corrugated ($\approx$0.8~{\AA} in our
work, similar to experimental estimates\cite{goler2013rta,lima2013ass}). The observed graphene interplanar distances near the interface are slightly
expanded compared to experimental bulk graphite (3.34~{\AA} \cite{baskin1955}) and in good
qualitative agreement with estimates from scanning tunneling microscopy
\cite{berger2010epo} and transmission electron microscopy.\cite{norimatsu2009}

We have also compared our findings to geometries for the straight PBE functional (no vdW
correction), and for the local-density approximation (LDA). In PBE,
the C-C interplanar distances are unphysically expanded 
(4.42~{\AA} for MLG).
In contrast, the LDA geometry of the carbon planes agrees qualitatively with PBE+vdW,
although LDA incorporates no long-range vdW interactions. The first qualitative
geometry difference between the PBE+vdW and LDA treatments appears in the Si part of the
top Si bilayer, where some Si atoms (those with dangling bonds pushing against the
$\pi$-bonded parts of the C interface plane \cite{varchon2007eso}) are located much deeper
(by $\approx$0.3~{\AA}) in PBE and PBE+vdW than in LDA. A direct crystallographic verification would be
desirable.

In a grand canonical formalism, the possible equilibrium conditions
for different surface phases can be represented by the chemical
potentials of C and Si, $\mu_\text{C}$ and $\mu_\text{Si}$. In
experiment, $\mu_\text{C}$ and $\mu_\text{Si}$ can be manipulated, e.g., 
through the substrate temperature and background pressure of gases
that supply Si or
C.\cite{reuter2002csa,tromp2009tak,heer2011laa,srivastava2012iso}. Consider
the surface energies $\gamma$ of a two-dimensional periodic SiC slab
with a C face and a Si face. In the limit of sufficiently thick slabs,
we have:   
\begin{equation}
  \gamma_\text{Si-face} + \gamma_\text{C-face} = \frac{1}{A} \left(
  E^\text{slab} - N_\text{Si} \mu_\text{Si} -
  N_\text{C} \mu_\text{C} \right) \, .
\end{equation}
$N_\text{Si}$ and $N_\text{C}$ denote the number of Si and C atoms in the slab,
respectively, and $A$ is the chosen area. All our surface energies are in eV per area of a
(1$\times$1) SiC unit cell. The letter $E$ denotes total energies for a given atomic
geometry throughout this work. In our actual calculations, we always choose a fixed
H-terminated C-face geometry, which cancels out for all surface energy differences related to
the Si face.

The major experimental ($T,p$) dependence during growth arises through
the reservoirs of Si and C, which define $\mu_\text{Si}$ and
$\mu_\text{C}$.\cite{reuter2002csa,tromp2009tak} Thus, a precise
control of background gases as reservoirs (for instance, disilane
\cite{tromp2009tak}) is desirable, even if 
calibration variations\cite{lu2010acr} may require exact ($T,p$)
ranges to be adjusted separately for a given growth chamber. The
actual growth process proceeds by Si out-diffusion from
underneath already formed graphene planes. Yet, the external Si
reservoir background pressure still matters in equilibrium: As long as the
diffusion path to the outside remains open, so does the inward
diffusion path, and near equilibrium with the reservoir gas can be
achieved. During intermediate stages of the formation of a new
graphene plane,\cite{norimatsu2009} such diffusion paths must be
available.  

In principle, we could further include the much smaller ($p,T$) dependence
of the solid phases by focusing on Gibbs free energies $G(T,p)$ in the
(quasi-)harmonic approximation instead of $E$. However,
quantifying this $T$ dependence precisely would here necessitate
accurate phonon calculations for structure sizes of the order of
$\approx$2,000~atoms, a task that is computationally prohibitive at
present. ZPC are small for the bulk phases (see SI
\cite{supplementary}). Still, the possible small contribution of
finite $T$ stresses at the growth conditions\cite{roehrl2008rso}
is kept in mind when interpreting our calculated results below. 

In equilibrium with a stable SiC bulk, 
$\mu_\text{C}$ and $\mu_\text{Si}$ are linked through
\begin{equation}
  \mu_\text{Si} + \mu_\text{C} = 2 E^\text{bulk}_\text{SiC} =
  E^\text{bulk}_\text{C} + E^\text{bulk}_\text{Si} + 2 \Delta
  H_f(\text{SiC}) \, .
\end{equation}
The energies are per atom, and $\Delta H_f(\text{SiC})$ is the
formation enthalpy of SiC with respect to the
elemental C and Si. The bulk phases define the chemical potential
limits within which the SiC crystal is stable against decomposition
into bulk Si or C: $\mu_\text{C}\le E^\text{bulk}_\text{C}$ and 
$\mu_\text{Si}\le E^\text{bulk}_\text{Si}$, leading to
\begin{equation}\label{Eq:limits}
  E^\text{bulk}_\text{C} + \Delta H_f(\text{SiC}) \le \mu_\text{C}
  \le E^\text{bulk}_\text{C} 
\end{equation}
and analogous for Si. The diamond structure for Si is the appropriate bulk phase, but for
C, there is a close competition between diamond and graphite.\cite{berman1955,yin1984} We
thus include both phases in our analysis.

\begin{figure*}[!htb]
  \includegraphics[width=1.0\textwidth]{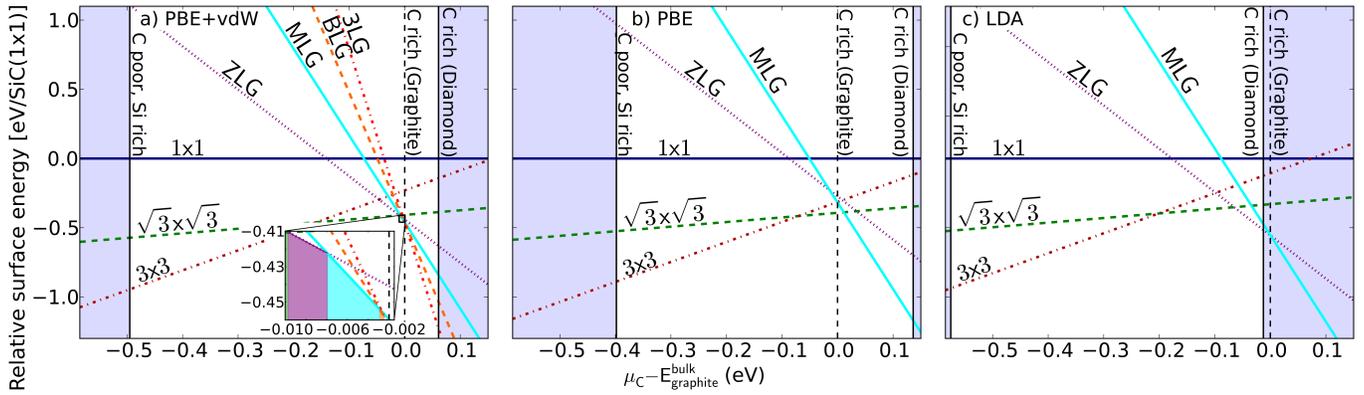}
  \caption{Comparison of the surface energies for five different reconstructions of the
3C-SiC(111) Si side, relative to the bulk-terminated (1$\times$1) phase (always unstable),
as a function of the C chemical potential within the allowed ranges (given by diamond Si,
diamond C or graphite C, respectively). (a) PBE+vdW, (b) PBE,
(c) LDA. The shaded areas indicate chemical potential values outside
the strict thermodynamic stability limits of Eq. \ref{Eq:limits}.}
\label{Fig:surf-energies}
\end{figure*}

The experimentally reported energy difference between diamond and
graphite at $T$=0~K is 25~meV/atom \cite{berman1955}. 
Based on the potential energy minima (no ZPC)
graphite is found to be more stable than diamond 
in PBE+vdW by 60~meV/atom (Fig.~\ref{Fig:surf-energies}a). This is
qualitatively consistent with the extrapolated 
experimental phase hierarchy. In plain PBE graphite is overstabilized 
by 130~meV/atom. In LDA, both phases are similarly stable: Considering
only the potential energy surface, diamond is slightly more stable (by
12~meV), but already the inclusion of ZPC\cite{yin1984} would
neutralize this balance (graphite more stable by 3~meV/atom\cite{florian2013ms}).

The surface energies of the known surface phases of SiC (Si face) are
shown as a function of $\Delta \mu_\text{C} = \mu_\text{C}-E^\text{bulk}_\text{C}$ in
Fig.~\ref{Fig:surf-energies}a for PBE+vdW. The most stable phase for a
given value of $\Delta \mu_\text{C}$ is that with the lowest surface
energy. Going through Fig.~\ref{Fig:surf-energies}a from left to right, we
find the expected broad ranges of stability for the Si-rich (3$\times$3) and
($\sqrt{3}\times\sqrt{3}$)-R30$^\circ$ phases. Just before the C-rich limit (bulk
graphite) is reached, there is a crossover first to ZLG, then to MLG, and even to a very
narrow slice of the BLG phase. As an additional bound, ABA-stacked trilayer graphene (3LG)
is also shown, crossing BLG within 1~meV of bulk graphite. 

While the respective stability ranges are narrow (inset of
Fig.~\ref{Fig:surf-energies}a: 4 meV, 5 meV and $<$1 meV for ZLG, MLG
and BLG, respectively at the chemical potential axis), but it is
important to recall that narrow chemical potential ranges do not
necessarily correspond to narrow experimental conditions: A drastic change in
the number of Si ($N_\text{Si}$) and C ($N_\text{C}$) atoms can correspond to a small
change of $\mu$. For instance, one would first have to remove all Si from the SiC crystal to
cross beyond the graphite stability line \emph{in equilibrium}. However, the
surface-energy differences between the different phases are also rather small
(a few tens of meV per (1$\times$1) SiC surface unit cell). The
primary approximations that we cannot systematically improve in our
calculations are the density functional used, as well as possible
small temperature-dependent surface strain effects (see
above). The key message of Fig.~\ref{Fig:surf-energies}a is thus that
MLG and its related phases all appear \emph{at least} as very near
equilibrium phases, a fact that is nonetheless critical for a
qualitatively correct understanding of their growth and properties.

What we can do is to show how our results would be affected by
different density-functional treatments. We have thus recomputed the
surface phase diagram up to the MLG phase for two widely used
functionals in Figs.~\ref{Fig:surf-energies}b and c: the plain PBE
functional, which lacks long-range vdW interactions and should thus yield
untrustworthy results, and for the often used LDA functional. As
expected, the absence of vdW tails in the plain PBE functional
changes the phase diagram drastically. Due to the overstabilization of
graphite (130~meV/atom), its stability line moves significantly further to
the left, as does the crossover point between ZLG and MLG.
As a result, \emph{neither} ZLG nor MLG becomes stable over the competing Si-rich
($\sqrt{3}\times\sqrt{3}$)-R30$^\circ$ phase in PBE, in outright contradiction to
experiment.\cite{tromp2009tak} In the LDA-derived phase diagram, the most significant
change compared to PBE+vdW is the apparent incorrect stability hierarchy of graphite vs.
diamond (without ZPC). If the diamond line were discounted, our calculations show the
ZLG-MLG crossover point almost exactly on the graphite line.
Still, even taking LDA at face value implies the existence of $T$-$p$ conditions very
close to equilibrium for MLG, making the experimental search for such conditions promising.

Figure~\ref{Fig:surf-energies} thus shows the most
important point of our paper: The existence of equilibrium or near-equilibrium chemical
potential ranges for ZLG, MLG, and even BLG, corresponding to specific $T$/$p$ conditions
in experiment. For each phase, this finding proves the potential for much better growth
control than what could be expected if each phase were just a necessary (but not
thermodynamically stable) kinetic intermediate. While true \emph{reversibility} for actual
MLG may be hard to achieve \cite{tromp2009tak,Hannonprivate} (the reverse growth process,
disassembling a fully formed graphene plane would be kinetically difficult), the active
\emph{forward} growth process from MLG to BLG under Si out-diffusion should still be
limitable by appropriate $T$/$p$ conditions. A macroscopically homogeneous surface very
close to pure-phase MLG should thus be achievable in principle.

Figure~\ref{Fig:surf-energies} shows unambiguously the importance of a consistently
accurate numerical treatment of the experimentally observed phases in their
large unit cells. It would obviously be much more economical to consider smaller-cell
approximant phases to the true
(6$\sqrt{3}\times$6$\sqrt{3}$)-R30$^\circ$ supercells. However, 
the residual artificial strain and inadequate bonding in those phases are too large
for meaningful surface energy comparisons.\cite{sclauzero2011lsi} For instance,
the popular ($\sqrt{3}\times\sqrt{3}$)-R30$^\circ$
\cite{varchon2007eso,mattausch2007ais} approximant would intersect the
graphite stability line at a surface energy
$\gamma|_{\mu_C=E^\text{bulk}_\text{graphite}}$=0.15~eV, far above the
actually stable phases. Likewise, a slightly rotated (5$\times$5)
approximant to the ZLG phase\cite{pankratov2010eso} (a periodicity
sometimes seen in experiment\cite{riedl2007spo,starke2009ego}) would intersect at
($\gamma|_{\mu_C=E^\text{bulk}_\text{graphite}}$=$-$0.35~eV), still
higher by 0.06~eV than even the closest competing Si-rich phase, the
($\sqrt{3}\times\sqrt{3}$)-R30$^\circ$ Si adatom phase
($\gamma|_{\mu_C=E^\text{bulk}_\text{graphite}}=-0.41$~eV). The (5$\times$5) phase is either a
nonequilibrium phase, or its structure is not the same as that assumed
in Ref. \cite{pankratov2010eso}. 

The true problem with artificially strained approximant phases is that
the resulting strain can obscure other electronically relevant
properties, such as the energetics of defects. As an example, we
consider a specific class of C-rich defects suggested as an
equilibrium feature of the ZLG phase in Ref.~\cite{qi2010ego}. Two
different defect positions, ``hollow'' and ``top'' were
suggested.\cite{qi2010ego} Indeed, both would be more stable than the
hypothetical ($\sqrt{3}\times\sqrt{3}$)-R30$^\circ$ ZLG approximant
when included there in a $(3\times 3)$ arrangement as done in
Ref.~\cite{qi2010ego}: $-$1.75~eV per defect for the hollow position,
$-$2.93~eV per defect for top, both at
$\mu_C=E^\text{bulk}_\text{graphite}$. However, the same defects are
unstable when included 
into and compared to the correct ($6\sqrt{3}\times
6\sqrt{3}$)-R30$^\circ$ ZLG phase: $+$5.28~eV per defect for hollow,
$+$5.27~eV for top, again at $\mu_C=E^\text{bulk}_\text{graphite}$
(see SI\cite{supplementary} for structure and other details). 

In conclusion, we can now rationalize some specific growth-related
observations: \\
(1) When simply heating a sample in UHV, the background pressures of Si and C are low and
ill-defined. The observed inferior morphologies and wide
variations of experimental conditions\cite{sutter2009,lu2010acr} are consistent with
this picture. \\
(2) Much more homogeneous growth can be achieved in an Ar atmosphere,\cite{emtsev2009tws}
although MLG/BLG phase areas still coexist. A uniform background partial pressure of
Si, however, will not be strictly guaranteed. \\
(3) The observed thermodynamic ZLG stability and improved growth of MLG
by controlling a disilane reservoir \cite{tromp2009tak} is fully consistent with our
findings. \\ 
(4) The use of a confined cavity to control the Si flux away from the sample reportedly
yields excellent wafer-size films.\cite{heer2011laa} Maintaining a controlled Si partial
pressure at constant temperature is most likely the important step.  \\
(5) Finally, similar considerations may also aid the much more difficult
growth of MLG on the C face.\cite{heer2011laa,hass2007spo} While our work is
restricted to the Si face, a well-defined interface layer on the C face at finite disilane
background pressure was proposed very recently.\cite{srivastava2012iso} This finding is an
excellent additional indication that near-equilibrium surface conditions are indeed the
key to the best possible epitaxial growth of graphene on SiC.



\end{document}